\documentclass[%
 reprint,
prb,
]{revtex4-2}

\usepackage{dcolumn}
\usepackage{amsmath}
\usepackage{float}
\usepackage{graphicx}
\usepackage{siunitx}
\usepackage[colorlinks=true, allcolors=blue]{hyperref}
\usepackage{bm}

\renewcommand{\eqref}[1]{Eq.~(\ref{eq:#1})}

\newcommand{\figref}[1]{Fig.~\ref{fig:#1}}
\newcommand{\Figref}[1]{Figure~\ref{fig:#1}}
\newcommand*{\vect}[1]{\bm{#1}}

\newcommand*{\Ev}{\vect{E}}
\newcommand*{\xv}{\vect{x}}

\begin{document}

\title{Tunable metasurface inverse design for \\80\% switching efficiencies and 144$^\circ$ angular steering}

\author{Haejun Chung}
\author{Owen D. Miller}%
\affiliation{Department of Applied Physics and Energy Sciences Institute, Yale University, New Haven, Connecticut 06511, USA}%





\begin{abstract}
Tunable metasurfaces have demonstrated the potential for dramatically enhanced functionality for applications including sensing, ranging and imaging. Liquid crystals (LCs) have fast switching speeds, low cost, and mature technological development, offering a versatile platform for electrical tunability. However, to date, electrically tunable metasurfaces are typically designed at a single operational state using physical intuition, without controlling alternate states and thus leading to limited switching efficiencies ($<30\%$) and small angular steering ($<\ang{25}$). Here, we use large-scale computational ``inverse design'' to discover high-performance designs through adjoint-based local-optimization design iterations within a global-optimization search. We study and explain the physics of these devices, which heavily rely on sophisticated resonator design to fully utilize the very small permittivity change incurred by switching the liquid-crystal voltage. The optimal devices show tunable steering angles ranging from \ang{12} to \ang{144} and switching efficiencies above 80\%, exhibiting 6X angular improvements and 6X efficiency improvements compared to the current state-of-the-art.
\end{abstract}

\maketitle

\section{Introduction}
\begin{figure*}[t]
\centering
\includegraphics[width=1.0\linewidth]{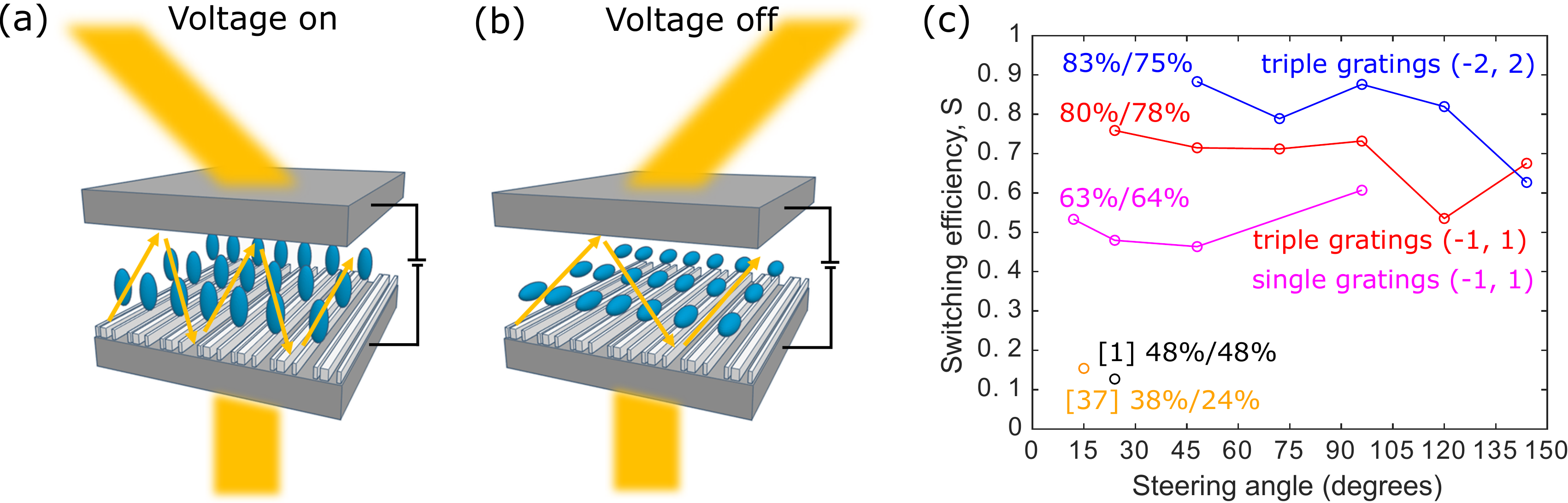}
\caption{Tunable metasurface via inverse design. (a) In the voltage-on state, the LC director is aligned along the vertical direction, perpendicular to TE-mode electric fields. A designed grating ideally steers incident light to the target angle. (b) When the voltage is off, the LC director is now parallel to TE-mode electric fields, thus maximizing the effective refractive-index change of the LC as seen by the fields. (c) Switching efficiency of our designed devices and existing devices over a wide range of steering angles. The efficiencies denoted within the figure (e.g. $83\%/75\%$) represent the highest diffraction efficiencies of each type of device in the on/off states. Switching efficiency, defined in \eqref{fom}, measures the distinguishability of the two operational states. The optimal devices show tunable steering angles ranging from \ang{12} to \ang{144}, offering 6X angular increase with the \ang{144} steering metasurface and 6X switching-efficiency improvement with the \ang{96} steering metasurface compared to the current state-of-the-art~\cite{li2019phase}.}
\label{fig:schematic}
\end{figure*}

Liquid-crystal (LC) devices promise the possibility for rapid electrical steering of optical beams, yet the complexity of designing for multiple refractive-index states in a single geometry has severely restricted the resulting diffraction efficiencies ($<50\%$), switching efficiencies (defined below, $<25\%$), and steering angles ($\leq \ang{24}$), even in state-of-the-art designs~\cite{li2019phase}. In this work, we use large-scale computational optimization, ``inverse design,'' to discover fabrication-ready designs with switching efficiencies and diffraction efficiencies above $80\%$ and steering angles up to \ang{144}. We combine adjoint-based gradients~\cite{Bendsoe2013,jensen2011topology,yang2009design,miller2012photonic,lalau2013adjoint,piggott2015inverse,molesky2018inverse,lin2018topology} for rapid local optimization within a larger global search to discover the high-efficiency and high-steering-angle designs. We compute the complex resonance patterns of the optimal devices, which reveal several competing design requirements that explain the need for computational optimization of many degrees of freedom. Unlike metasurfaces designed for lens-like focusing~\cite{aieta2015multiwavelength,khorasaninejad2015achromatic,avayu2017composite,shrestha2018broadband, chen2017gan,paniagua2018metalens,chen2018broadband,wang2018broadband} and related applications~\cite{zheng2015metasurface,ni2013metasurface,zhao2016full,arbabi2017planar}, we find that the optimal devices should have their field intensities concentrated not in the high-index grating material but instead in the low-index liquid-crystal embedding medium, to enable high switching efficiencies even for the relatively small refractive-index changes of LCs. Our largest-steering-angle devices exhibit $90\%$ diffraction efficiency at $-\ang{72}$ in the off state, and $70\%$ diffraction efficiency at $+\ang{72}$ in the on state, simultaneously exhibiting 6X angular and almost 6X switching-efficiency enhancements over the current state-of-the-art, paving a pathway to efficient liquid-crystal beam-control devices for applications ranging from LIDAR~\cite{poulton2019long,poulton2017coherent} to spatial light modulator~\cite{shrestha2015high}. 

Thin optical films with complex lithographic patterns can control phases, amplitudes, diffraction-order excitations, and more general wave dynamics with high efficiency over large-area devices, comprising the basis for the emerging field of metasurfaces~\cite{yu2014flat,aieta2015multiwavelength}. Metasurfaces have shown significant promise for static (non-tunable) applications such as holography~\cite{zheng2015metasurface, ni2013metasurface,zhao2016full}, lensing~\cite{aieta2015multiwavelength,khorasaninejad2015achromatic,avayu2017composite,shrestha2018broadband,chen2017gan,paniagua2018metalens,chen2018broadband,wang2018broadband}, and beam converters~\cite{wu2017broadband,chong2015polarization}, in large part due to the use of a relatively simple design principle: for a given frequency of interest, one can specify the desired outgoing phases and amplitudes (and possibly dispersion characteristics~\cite{yu2014flat}) across the device surface, and select from a library of waveguide-like meta-elements to locally approximate those phases and amplitudes. This design principle is not exact---the local-periodicity assumption is a source of error, especially in high-NA lens applications~\cite{chung2019high,lin2019topology}---and there is significant effort to leverage computation to improve it ~\cite{sell2017periodic,pestourie2018inverse,lin2019topology}, but it has been sufficient for proof-of-principle high-performance devices.

For \emph{dynamic} applications, however, in which the properties of the metasurface are designed to offer varying functionality in multiple operational states, from electrical~\cite{huang2016gate,yao2014electrically,sautter2015active,holsteen2019temporal}, mechanical~\cite{ee2016tunable}, or thermal~\cite{komar2018dynamic} switching mechanisms, the simple design principle appears to be quite inefficient. One might imagine that multi-state operation would require only small extra considerations in the ``library'' of designs, accounting for the additional states. However, as we show below, the requirement for high-efficiency resonant behavior in multiple states rapidly leads to highly complex resonant patterns, with individual elements far more complex than those of typical metasurface applications, due to the requirement for the multi-state behavior to be supported by a single geometrical structure. In lieu of a multi-state design principle, previous approaches~\cite{savo2014liquid,yao2014electrically,lee2014ultrafast,komar2017electrically} have simplified the design process by focusing only on high efficiency for a single state, but this naturally leads to lower efficiencies in the switching process over the dynamic range of the devices. (An alternative approach is to use a frequency comb in tandem with a designed metasurface, which can create time-dependent beam profiles albeit without full temporal ``steering'' control~\cite{shaltout2019spatiotemporal}.)

In this work, we show that large-scale computational design, an approach that efficiently optimizes over arbitrarily many degrees of freedom, offers a pathway to high-efficiency dynamic (tunable) metasurfaces. We focus on beam-switching with liquid-crystal devices, which already have significant commercial development and which show promise for applications such as LIDAR. We discuss the complexity of the design space, and describe a combined application of adjoint-based local-optimization techniques within a larger global-optimization platform, and use this approach to discover two-state switching devices with high switching efficiencies and high steering angles.


\section{Computational multi-state design}
\Figref{schematic}(a,b) is a schematic depiction of a liquid-crystal (LC) beam-switching device. As is typical in LC devices~\cite{bohn2018active,komar2017electrically}, the liquid-crystal layer is embedded between two alignment and contact layers. Within the liquid-crystal region, and above and below the contact layers, complex patterns can be lithographically fabricated, and previous work has designed grating layers for moderate-efficiency electrical~\cite{li2019phase, komar2017electrically,buchnev2015electrically} and thermal~\cite{komar2018dynamic} switching of LC devices. The key metric to design for is the \emph{switching efficiency}, i.e. how effectively the device can switch between different optical-beam patterns. For periodic grating and meta-grating structures, diffraction efficiency is an important determinant of the switching efficiency, but not the only one: a device that separates an optical field into a 50\% mix of two outgoing diffraction orders, for both voltages of a two-state device, effectively has zero switching efficiency due to the inability to distinguish the two states. Moreover, in many cases back-reflected light represents only a minor loss mechanism, without affecting the \emph{relative} power distribution between the forward-going beams nor the ability to distinguish them, and can be normalized out. Thus, for a two-state optical-switching device operating over frequencies $\omega$ with geometrical degrees of freedom $g$, we define switching efficiency by the expression
\begin{equation}
    S = \frac{1}{T} \left\{ \frac{1}{2} \sum_{s=\textrm{on,off}}\left[P^s_\textrm{tar}(\omega,g) - \sum_{j\neq \textrm{tar}}P^s_j(\omega,g) \right]\right\},
    \label{eq:fom}
\end{equation}
which is the power in the target (desired) diffraction orders, $P^{s}_{\rm tar}(\omega,g)$, averaged over state $s$, minus the total state-averaged power in all other diffraction orders, $P_{j\neq \textrm{tar}}^s(\omega,g)$, normalized by the forward transmission efficiency $T$. This definition of switching efficiency, which can be easily generalized to more states, linear combinations of diffraction orders, etc., enables comparison among different device designs. \Figref{schematic}(c) shows the switching efficiencies of recent state-of-the-art LC beam-switching devices, which show moderate diffraction efficiencies (24--54\%, labeled), but somewhat lower switching efficiencies (ranging from $13\%$ to $29\%$)) due to the contamination of unwanted diffraction orders that inhibits the ability to distinguish the on/off states. Included in \figref{schematic}(c) are the switching efficiencies of the optimal devices that we discover, discussed further below, segregated into three architectures: a class of devices with a single silicon grating in the liquid-crystal region (solid purple line), and two classes of devices with two additional gratings on top and bottom, one for -1 to +1 order steering (solid red line), and one for -2 to +2 order steering (solid blue line). There are many geometrical degrees of freedom in each architecture: the individual``pixels'' (78 nm wide) of each grating, the thicknesses of the alignment, contact, and liquid-crystal layers, and the period of the structure. The pixel size is chosen as $\lambda/20$ to provide sufficient control while avoiding features that are too fine for fabrication.
We take the switching to occur between two states with the same polarization, in which case the gratings can be chosen to have translation invariance perpendicular to a plane containing both angles, and the system can be modeled in this two-dimensional plane. There are many grating degrees of freedom ($\approx 400$), and to optimize these it is critical to be able to rapidly compute gradients of the switching-efficiency objective. To do so, we use the adjoint method (also known as ``topology optimization''~\cite{bendsoe2001topology,Bendsoe2013} and ``inverse design'' ~\cite{yang2009design,jensen2011topology,frandsen2014topology, lalau2013adjoint,piggott2015inverse,su2017inverse, sell2017large,callewaert2018inverse, ganapati2014light} in nanophotonics, and ``backpropagation'' in the deep-learning community~\cite{Werbos1994,Rumelhart1986,LeCun1989}), which is efficient and effective at optimizing many small-scale degrees of freedom~\cite{molesky2018inverse}. Adjoint-based methods exploit reciprocity (or generalized reciprocity~\cite{miller2012photonic}) to convert the process of computing thousands or millions of individual gradient calculations into a single extra simulation, in which ``adjoint sources'' are specified according to the desired objective, back-propagated through the optical system, and then combined with the ``direct'' fields excited by the original incident wave to compute all gradients at once. For an objective such as switching efficiency, \eqref{fom}, that depends on the outgoing electric fields $\Ev$, the general prescription~\cite{miller2012photonic} for each ``forward'' simulation (in this case, the voltage-on and voltage-off simulations) is to run an ``adjoint'' simulation with current sources proportional to the derivative of the objective with respect to the electric field (SM):
\begin{align}
    \vect{J}_{\rm adj}(\xv) &= -i\omega \frac{\partial \mathcal{F}}{\partial \Ev} \nonumber \\
                            &= -\frac{i\omega}{2}[c_\textrm{tar} \cos\theta_{\rm tar} \vect{E}_\textrm{tar}^{*}(\xv) - \sum_{n \neq \textrm{tar}}c_\textrm{n} \cos\theta_n \vect{E}_\textrm{n}^{*}(\xv)].
                            \label{eq:Jadj}
\end{align}
In our adjoint equation indicated by \eqref{Jadj}, we exclude the $1/T$ of \eqref{fom} to drive the optimization to exhibit high transmission in addition to high switching efficiency. The pixels in the gratings are represented during the design process as grayscale pixels, with refractive indices varying between their minimum and maximum values, and as the local optimization proceeds we penalize intermediate refractive-index values until a binary design is reached. This process is very efficient for the many grating degrees of freedom. However, it is less efficient for variables representing larger geometrical parameters: the thicknesses of the various regions, and the periodicity of the structure. Wave-interference effects create a tremendous number of poor-quality, local optima for these parameters, since varying them even by half a wavelength or less can take one from a field minimum to a maximum.

The many-local-optima problems for these ``global'' (beyond wavelength-scale) parameters could be significantly compensated by separating them into pixelated local degrees of freedom (DOFs) that vary independently. However, they are fixed by fabrication constraints and must not be separated. Thus, to optimize these parameters, we embedded the grating-DOF local-optimization procedure into a global search to discover optimal thickness and periodicity values. Particle swarm optimization~\cite{robinson2004particle} is used for a global optimization algorithm, initially instantiating many ``particles'' with random structural parameters (i.e., top TiO$_2$ grating, ITO, alignment, LC, silicon grating and bottom TiO$_2$ grating thicknesses). Within Each ``particle'' we perform inverse design, computed in a single computational core, optimizing the fine-scale features of the device. Then, new parameters of the next iteration are determined by the optimal figure of merit values. The global optimization is run for 150 iterations, which is sufficient to converge to a set of very similar ``particles,'' with similar large-scale-feature values. Each iteration of global optimization takes approximately 10 minutes on 25 cores in our computational cluster (Intel Xeon E5-2660 v4 3.2 GHz processors) while each inverse design iteration takes less than 5 seconds in a single core computer.

\section{Optimal Designs}
\begin{figure*}[t]
\centering
\includegraphics[width=0.8\linewidth]{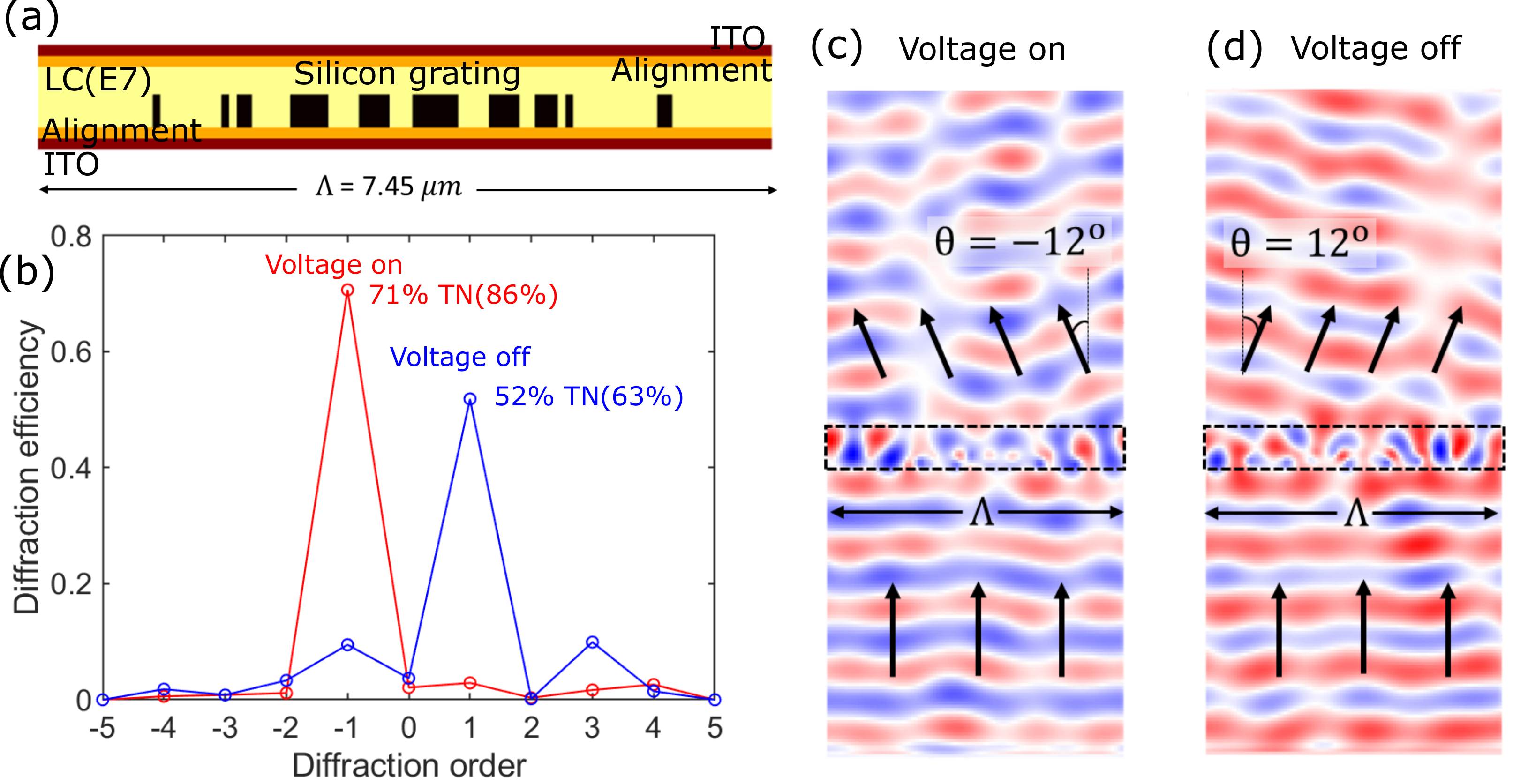}
\caption{Optimization of a single-grating tunable metasurface via inverse design at \SI{1550}{nm} wavelength. (a) Optimized structure showing silicon grating inside the LC-layer (E7). The period of the structure is 4.81$\lambda$-long (7.45 $\mu\textrm{m}$). The LC region and silicon grating have  thicknesses of \SI{698}{nm} and \SI{388}{nm}, respectively. The ITO and alightment layers have \SI{77}{nm} thickness. (b) Diffraction efficiency of the optimized device at \SI{1550}{nm} wavelength. The unwanted order diffractions are suppressed well via the FOM defined in this work, maintaining high efficiencies in the target orders. (c,d) Real parts of the electric field for the (c) voltage-on and (d) voltage-off states. They show clear outgoing fields propagating towards $\pm\ang{12}$.}
\label{fig:single_layer}
\end{figure*}

\begin{figure*}[htbp]
\centering
\includegraphics[width=0.9\linewidth]{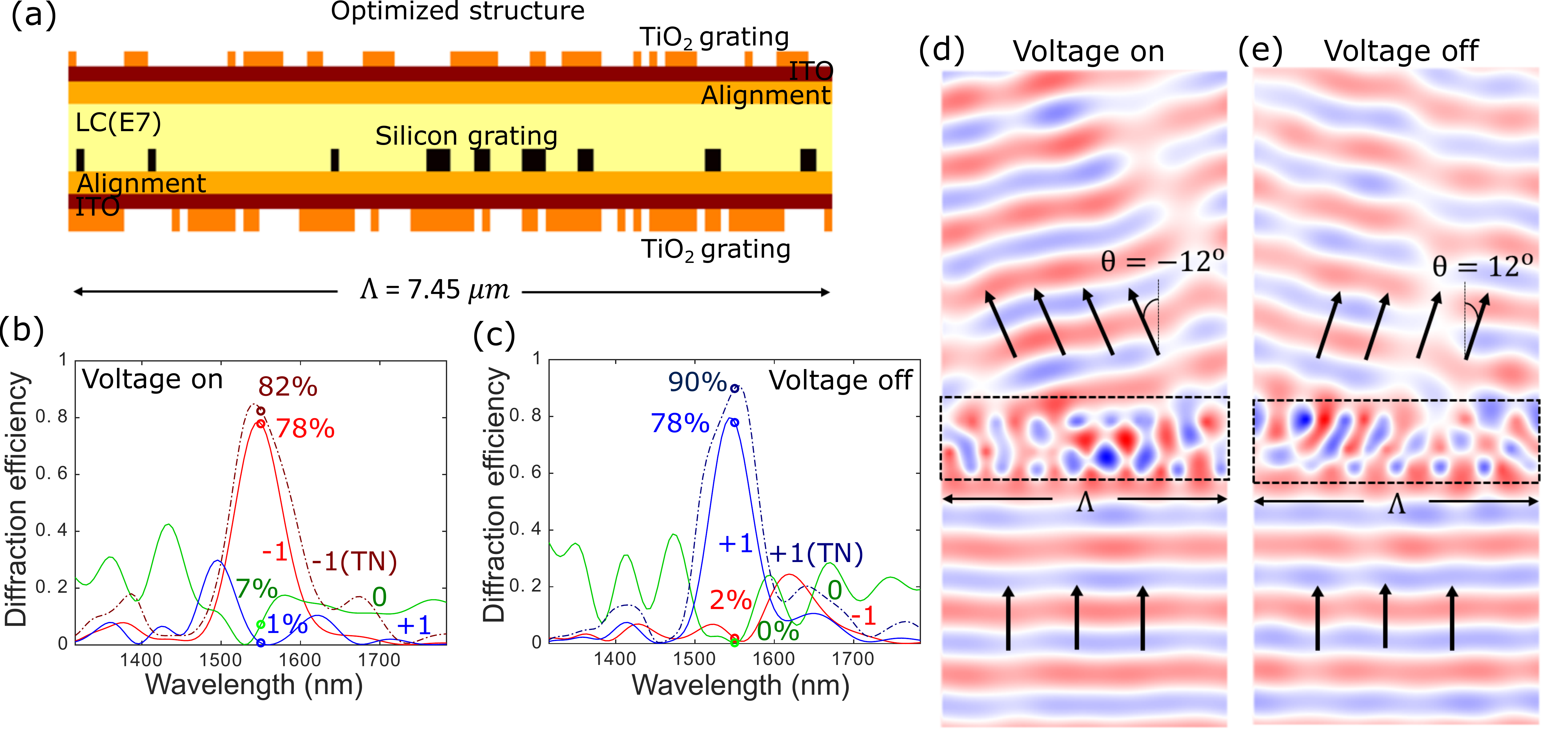}
\caption{Optimized triple-grating tunable metasurface for narrow-angle ($\pm\ang{12}$) steering with high efficiency. (a) Optimized structure showing triple gratings (TiO$_2$/Si/TiO$_2$). The period of the structure is 4.81$\lambda$ (7.45 $\mu\textrm{m}$), and the thicknesses of the LC, silicon grating, top TiO$_2$ and bottom TiO$_2$ are \SI{543}{nm}, \SI{155}{nm}, \SI{155}{nm}, and \SI{233}{nm}, respectively. The ITO is \SI{77}{nm} thick and the alignment layer is \SI{155}{nm} thick. (b,c) Diffraction efficiencies over infrared wavelengths for the (b) voltage-on and (c) voltage-off states. ``TN'' denotes transmission normalized efficiency. The optimized grating shows -1 order efficiency of 78\% in the voltage-on state and +1 order efficiency of 78\% in the voltage-off state. The transmission-normalized efficiencies are 82\% and 90\%, respectively. (d,e) Real parts of the electric fields for the (d) voltage-on and (e) voltage-off states.}
\label{fig:narrow_angle}

\bigskip
\includegraphics[width=1.0\linewidth]{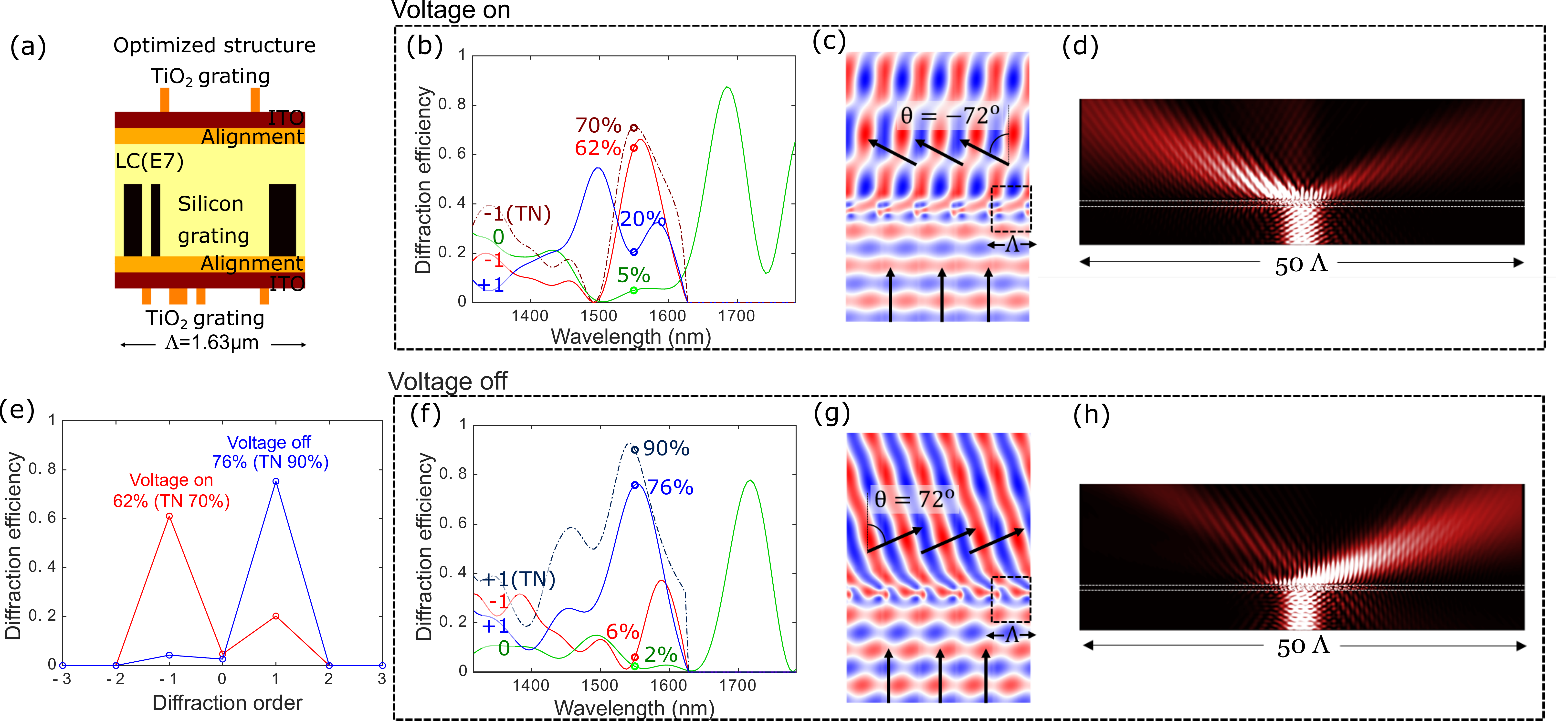}
\caption{Optimized triple-grating tunable metasurface for ultra-wide-angle ($\pm\ang{72}$) steering. (a) Optimized structure showing triple gratings (TiO$_2$/Si/TiO$_2$). The period of the structure is 1.05$\lambda$ (1.63 $\mu\textrm{m}$), and the thicknesses of the LC, silicon grating, top TiO$_2$ and bottom TiO$_2$ are \SI{930}{nm}, \SI{620}{nm}, \SI{233}{nm}, and \SI{155}{nm}, respectively. The ITO and the alignment layer both have \SI{77}{nm} thicknesses. (b,f) Diffraction efficiencies over the infrared spectrum for the (b) voltage-on and (f) voltage-off states. TN means transmission normalized efficiency. The optimized grating shows -1 order diffraction efficiency of 62\% in the voltage-on state and +1 order efficiency of 76\% in the voltage-off state. The transmission-normalized efficiencies are 70\% and 90\%, respectively. (c,g) Real parts of the electric fields for the (c) voltage-on and (g) voltage-off states. (d,h) E-field intensity profile at \SI{1550}{nm} wavelength for a supercell with an array of 50 unit cells excited by a Gaussian beam of 5-$\lambda$-width centered at \SI{1550}{nm} wavelength. The white dashed lines indicate the areas of the supercells.}
\label{fig:wide_angle}
\end{figure*}

We apply the multi-state computational design process described above to discover the single- and multi-grating designs depicted in Figs.~\ref{fig:single_layer}--\ref{fig:wide_angle}. We start by designing LC metasurfaces with a single embedded silicon grating, intentionally selecting a platform very similar to that of recent works~\cite{li2019phase,komar2018dynamic} to show the efficiency gains that are possible through computational design. Then we expand to structures with multiple grating layers, where we show the extensive capability for LC metasurfaces to simultaneously achieve high efficiency and high steering angles. In all of the designs demonstrated below,  we use \SI{1550}{nm} as our design wavelength. For the LC material, we use E7~\cite{yang2010complex}, which has a refractive-index variation $\Delta n$ of about 0.192 between the voltage-on and voltage-off states. TiO$_2$~\cite{palik1998handbook} is used for top and bottom supportive gratings while we use silicon for the grating inside the LC layer. ITO~\cite{laux1998room} and alignment layers~\cite{ma2018liquid,bohn2018active,komar2018dynamic} are included, as typically required. Unlike metasurfaces for lensing and related applications, high-index materials do not appear to be required for high diffraction efficiency nor switching efficiency; we use Si and TiO$_2$ simply because of their common usage~\cite{li2019phase, komar2017electrically,komar2018dynamic} and scale-up feasibility. The top ITO works as an electrical contact and the alignment layer coordinates the axis of the LC director into the out-of-plane direction. Of course, a different wavelength, set of materials, or parameter regime can seamlessly be incorporated into our design process.

\subsection{Single-grating designs}
In this section, we design tunable metasurfaces with a single grating layer. Single-grating metasurfaces can be designed by physical intuition using effective medium theory~\cite{choy2015effective}, whereby the filling ratio of two materials is adjusted to realize specific transmission phases, or by a unit-cell library approach~\cite{aieta2015multiwavelength}, whereby a large design space is decomposed into ``unit cells'' with a small number of parameters whose entire design space can be stored in a library to design for a small number of criteria. Neither approach is well-suited to designing many parameters for multi-state operation.

\Figref{single_layer} shows an optimal single-grating design for switching between $-\ang{12}$ and $+\ang{12}$, angles chosen to match the current state-of-the-art~\cite{li2019phase}. The optimized single-grating metasurface achieves diffraction efficiencies of 71\% in the voltage-on state and 52\% in the voltage-off state, with clean outgoing field patterns visible in Figs.~\ref{fig:single_layer}(c,d). A key determinant of the angular purity is the diffraction efficiency normalized by the total transmission, since reflection does not contribute noise in the outgoing-wave patterns, and the transmission-normalized (TN) efficiencies of this structures are 86\% and 63\%, respectively. During the optimization, we fix the top and bottom sides have to have thin ITO and alignment layers, while we include the thicknesses of the LC layer and the silicon grating as global design parameters. The beam steering efficiency (48\%) shown here is already significantly larger than any other theoretical designs. However, for key applications, one can expect the need for larger steering angles and even high switching efficiencies. Thus, in the next section we explore more complex device architectures.


\begin{figure*}[t]
\centering
\includegraphics[width=0.8\linewidth]{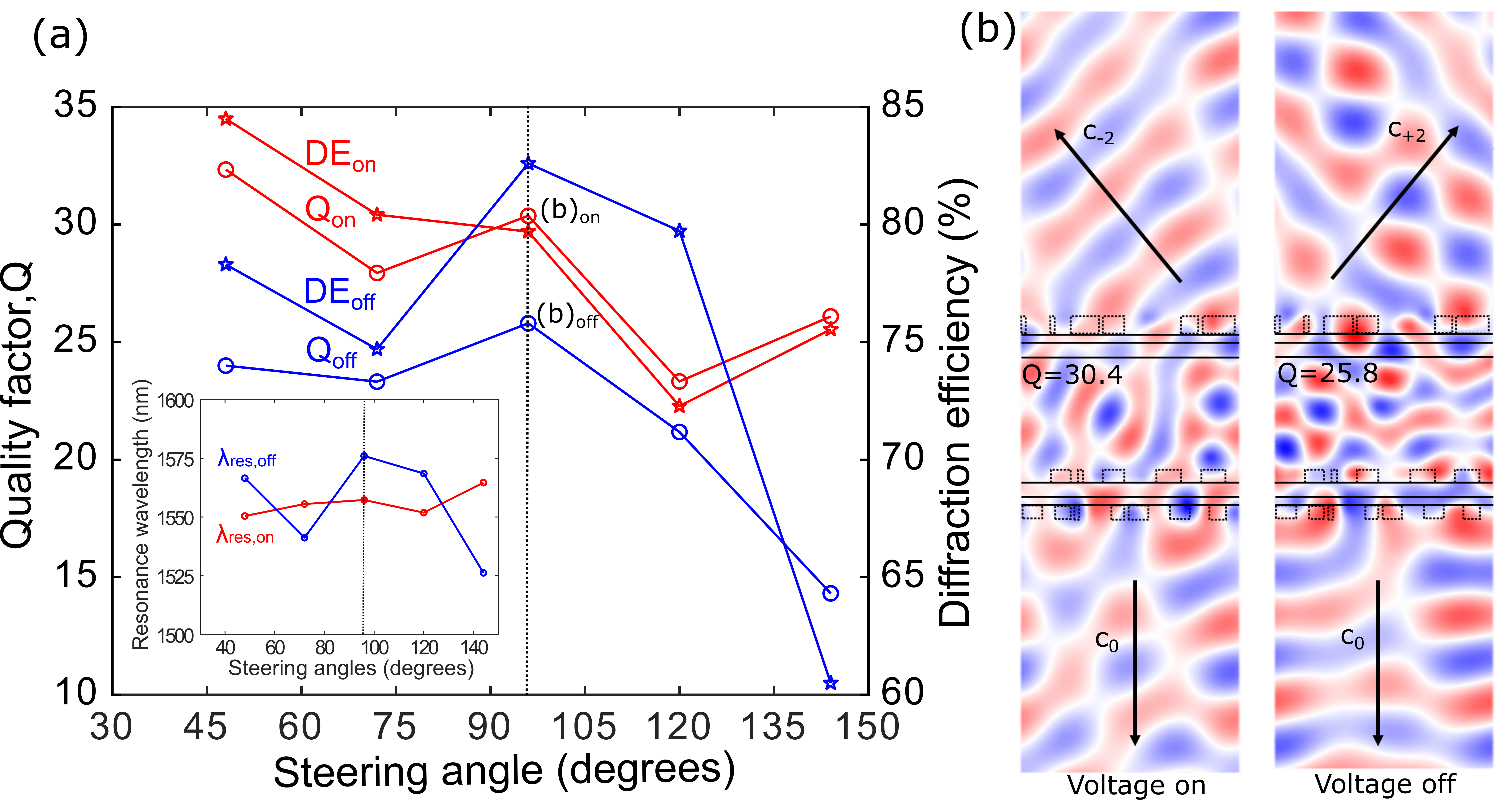}
\caption{Resonance analysis of the beam-steering metasurfaces (a) Quality factor and diffraction efficiency versus steering angle for the triple-grating metasurfaces (-2 to +2). There is a strong correlation between the quality factors of the devices (circles) and their diffraction efficiencies (stars). The inset figure shows the resonance wavelengths of the major resonance patterns found in the optimum devices. (b) Real parts of the electric fields at their resonant frequencies (\SI{1571}{nm} for the voltage-on state and \SI{1550}{nm} for the voltage-off state) for \ang{96} steering angle [black dashed lines in (a)]. Quality factors found in this structure are 30.4 (on) and 25.8 (off) and the resonances strongly couple to the incident and target channels.  }
\label{fig:resonant_mode}
\end{figure*}

\subsection{Multi-grating designs}
In this section, we design tunable metasurfaces with three gratings---one in the silicon, and two in the TiO$_2$ surroundings---to discover an ultra-high-efficiency beam-switching device. Generally, multi-layer metasurface structures offer increased functionality through increased path-length enhancements and multiple-reflection interactions, and multilayer metasurfaces have been proposed for light concentration~\cite{lin2019topology} and flat lens~\cite{lin2018topology} applications. Here, the two TiO$_2$ gratings must enable specific functionality: the bottom grating must be transmissive for the plane wave incident from below, while being highly reflective for all off-angle plane waves reflected from above, and the top grating should either redirect all light to the single desired outgoing diffraction order, or at least restrict transmission through any undesired orders. Though the use of multiple gratings requires precise alignment, the gratings play complementary roles and potentially enable near-unity switching efficiencies even at very high switching angles.

We start by reconsidering the problem of high-efficiency switching from $-\ang{12}$ to $+\ang{12}$. The optimal design, depicted in \figref{narrow_angle}(a), achieves diffraction efficiencies of 78\% (82\% TN) in the voltage-on state and 78\% (90\% TN) in the voltage-off state, for a switching efficiency of 76\%, with very little power in any other outgoing diffraction orders. The clean outgoing waves are depicted in \figref{narrow_angle}(c,d). The optimized device has ITO/alignment layer thicknesses of 77, 155 nm, top and bottom grating thicknesses of 155 and 233 nm, a silicon grating thickness of 155 nm, and a liquid-crystal layer thickness of 543 nm. 

Among the many designs that were discovered across the single-, double-, and triple-grating architectures, for beam-steering angles from $\ang{24}$ to $\ang{144}$, we highlight here the highest steering-angle designs, which employ a triple-grating structure to achieve steering from $-\ang{72}$ to $+\ang{72}$. By avoiding a design with collections of locally varying ``unit cells,'' we circumvent the limitations~\cite{chung2019high} arising from breaking the local-periodicity assumption. \Figref{wide_angle} shows an optimal structure with thickness \SI{1.684}{\mu m} and period \SI{1.63}{\mu m} (which is $1.05\lambda$). The diffraction efficiencies in the target orders are 62\% and 76\% for voltage-on and -off states, respectively, while the transmission normalized target efficiencies are 70\% and 90\%, respectively. These diffraction efficiencies are individually nearly as large as those of state-of-the-art high-angle diffraction gratings that are designed only for a single operational state~\cite{sell2017large}. The real part of the electric-field profile shown in \figref{wide_angle}(c),(d) demonstrates the clear angle-directed outgoing wave patterns, and in \figref{wide_angle}(d),(h) we simulate a Gaussian beam incident upon the structure to more clearly visualize the high-fidelity switching that is achieved.

To understand the physics of the high-efficiency designs that we discover, we analyze the quality factors ($Q$) and resonance patterns for our high-efficiency structures designed for -2 to +2 order switching. As shown in \figref{resonant_mode}(a), we find that the optimal designs are a new kind of \emph{dual-resonance} structure that support one moderate-$Q$ resonance in the on state (red circles), and a different moderate-$Q$ resonance in the off-state (blue circles). In \figref{resonant_mode}(b), we depict the resonant field pattern of our $\ang{96}$-steering device, which has 87\% switching efficiency and transmission-normalized efficiencies of 86\% (voltage-on) and 95\% (voltage-off). The field pattern is computed by exciting point dipoles at the high-intensity locations of the plane-wave forward simulation, discovering the mode responsible for the high efficiency. The resonance pattern of the voltage-on state shows strong coupling to $c_{-2}$ channel in transmission direction and $c_{0}$ channel in incidence direction, agreeing well with what one would expect. In the voltage-off state, the new resonant pattern couples to the $c_{+2}$ channel in transmission direction and $c_{0}$ channel in incidence direction. An intriguing trend in \figref{resonant_mode}(a) is that when we overlay the diffraction efficiencies (red and blue stars) with the quality factors, we observe a correlation between the two. This suggests that quality factors of at least 30 or so may be necessary to achieve the highest possible diffraction efficiencies in each operational state of a beam-switching or beam-steering device.

\section{Extensions}
In this work, we have demonstrated high-efficiency, wide-angle, electrically tunable metasurfaces that operate at \SI{1550}{nm} wavelength, achieving state-of-the-art steering angles and switching efficiencies. Our inverse-design approach can be applied more broadly to any multi-configuration-state optical functionality, for applications including next-generation LiDAR, spatial light modulators, and free-space data communication. In the liquid-crystal beam-steering design space, natural extensions include many-state operation (towards steering rather than switching) and three-dimensional beam control. In addition to the ``bottom-up'' large-scale optimization approach presented here, an interesting question is the limits of such design: for a given set of liquid-crystal and semiconductor refractive indices, is it possible to exploit sum rules~\cite{gordon_1963,purcell_1969,sohl_gustafsson_kristensson_2007,sanders_manjavacas_2018,Shim2019}, passivity and convexity~\cite{kwon_pozar_2009,liberal_ederra_gonzalo_ziolkowski_2014,miller_polimeridis_reid_hsu_delacy_joannopoulos_soljacic_johnson_2016,Zhang2019}, and/or duality~\cite{angeris2019computational} to map out the limits to maximal performance as a function of the steering angle and the number of operational states?

\section*{Funding Information}
H.~C. and O. D. M. were partially supported by the Air Force Office of Scientific Research under award number FA9550-17-1-0093.

%

\end{document}